\documentclass[numberedappendix,appendixfloats]{emulateapj}
\usepackage{natbib}

\shortauthors{T. Hosokawa et al.}
\shorttitle{Stellar Ages from Pre-Main Sequence Evolution}

\newcommand{\msun}{M_{\odot}}
\newcommand{\msunyr}{M_\odot~{\rm yr}^{-1}}

\begin{document}

\title{On the Reliability of Stellar Ages and Age Spreads Inferred from 
Pre-Main Sequence Evolutionary Models}

\author{Takashi Hosokawa\altaffilmark{1,2},
       Stella S.R. Offner\altaffilmark{3},
       Mark R. Krumholz\altaffilmark{4}}

\altaffiltext{1}{Jet Propulsion Laboratory, California Institute   
of Technology, Pasadena CA 91109, USA; 
Takashi.Hosokawa@jpl.nasa.gov, hosokwtk@gmail.com} 
\altaffiltext{2}{Department of Physics, Kyoto University, 
Kyoto 606-8502, Japan}
\altaffiltext{3}{Harvard-Smithsonian Center for Astrophysics, 60 Garden
Street, Cambridge, MA 02138, USA}
\altaffiltext{4}{Department of Astronomy and Astrophysics, University of
California, Santa Cruz, CA, 95064, USA}

\begin{abstract}
We revisit the problem of low-mass pre-main-sequence (PMS)
stellar evolution and its observational consequences for where stars 
fall on the Hertzsprung-Russell diagram (HRD). 
In contrast to most previous work,
our models follow stars as they grow from small masses via accretion,
and we perform a systematic study of how the stars' HRD evolution is
influenced by their initial radius, by the radiative properties of the
accretion flow, and by the accretion history, using both simple
idealized accretion histories and histories taken from numerical
simulations of star cluster formation. We compare our numerical
results to both non-accreting isochrones and to the positions of
observed stars in the HRD, with a goal of determining whether both the
absolute ages and the age dispersions inferred from non-accreting
isochrones are reliable. 
We show that non-accreting isochrones can
sometimes overestimate stellar ages for more massive stars (those with 
effective temperatures above $\sim 3500$ K), thereby
explaining why non-accreting isochrones often suggest a systematic 
age difference between more and less massive stars in the same cluster. 
However, we also find the only way to produce a similar overestimate for
the ages of cooler stars is 
if these stars grow from $\sim 0.01$ $\msun$ seed 
protostars that are an order of magnitude smaller than predicted by
current theoretical models, and if the size of the seed protostar
correlates systematically with the final stellar mass at the end of
accretion. We therefore conclude that, unless both of these conditions
are met, inferred ages and age spreads for cool
stars are reliable, at least to the extent that 
the observed bolometric luminosities and temperatures are accurate.
Finally, we note that the time-dependence of the mass accretion rate
has remarkably little effect on low-mass stars' evolution on the HRD,
and that such time-dependence may be neglected for all stars except 
those with effective temperatures above $\sim 4000$ K.
\end{abstract}


\section{Introduction}
\label{sec:intro}

Pre-main-sequence (PMS) stars in low-mass star forming regions
show a sizable luminosity spread when placed on 
the Hertzsprung-Russell diagram (HRD) \citep[e.g.,][]{hillenbrand09}.
This spread translates into a significant dispersion in inferred 
stellar ages that records the past star formation 
activity in each region
(e.g., \citealt{dantona94, baraffe98, siess00, palla00, palla99, 
hartmann01,hartmann03}).
However, the idea that star clusters form over an extended period is subject to
extensive debate on both observational and theoretical grounds
\citep[e.g.,][]{elmegreen00a, hartmann01a, tan06a, krumholz07a, evans09a},
and several authors have claimed that the dispersion of stellar
luminosities does not reflect a real age spread.
Members of young binaries and multiples exhibit a tighter age
correlation, supporting the existence of an intrinsic age distribution.  
However, a luminosity spread persists even among such systems, 
and some companions display a substantial age mismatch
\citep{prato03, stassun08,krauss09}. 
Deriving stellar ages is complicated from an observational 
standpoint. 
Calculation of stellar bolometric luminosities is
beset by uncertainties in extinction, photometric variability, and
unresolved multiplicity. 
Calibration between the stellar spectral type and effective 
temperature is also not trivial.
In some cases, it can be demonstrated that 
observational uncertainties alone are sufficient to induce an age spread of
$> 10$ Myr and mask a coeval stellar population \citep{slesnick08}. 
However, \citet{dario10b, dario10a} carefully model these
uncertainties and conclude that these effects alone cannot reproduce  
the entire spread.
Other age indicators such as stellar rotation rate
\citep{jeffries10}, surface gravity \citep{slesnick08}, and lithium
abundances \citep{sestito08} also support the idea that the inferred age
spreads are real, but are each subject to significant challenges.

Apart from the observational uncertainties, physical mechanisms may be 
responsible for a portion of the observed HRD scatter.
For the purpose of inferring stellar ages, it is usually assumed that PMS stars 
first appear along a ``birthline'' in the HRD 
when mass accretion ceases (e.g., \citealt{PS90,hartmann97}). 
However, luminosities of younger embedded stars (Class 0 and I sources)
that are presumably still accreting
also show a wide spread, and a fraction of them have luminosities much lower than 
the values expected from the standard birthline 
(e.g., \citealt{kenyon90, evans09, enoch09}).
A solution for this ``luminosity problem'' is the scenario that 
mass accretion takes place very time-dependently, 
repeating burst-like accretion phases and quiescent phases. 
Recent numerical simulations suggest that such episodic
mass accretion is caused by gravitational fragmentation
of a circmustellar disk (e.g., \citealt{vorobyov05, machida11}),
though radiative warming from protostars alleviates
it \citep[e.g.,][]{offner09}. Regardless of the ultimate explanation for the
luminosities of Class 0 and I sources, the existence of young stars that
fall well below the putative birth line is strong evidence that we must
extend our PMS evolution models to include the accretion phase.

\citet[][hereafter BCG09]{baraffe09} study  
protostellar evolution with various episodic mass accretion histories
and examine the resultant spread of PMS stars in the HRD.
They argue that PMS stars of the same mass and age show 
some scatter in the HRD owing to variation of
the early evolution resulting from complex accretion histories.
However, BCG09 simultaneously vary not only their accretion histories,
but also their initial stellar models and the radiative properties of
the accretion flow. 
Because they change these parameters in
correlated ways and do not perform a systematic survey of parameter
space, it is not clear which of these effects drives their results. Nor is
it clear whether the results they generate via their parameter choices
are consistent with observed HRDs of clusters. Consequently, it is still
unclear how much vigorous time-dependent accretion histories
influence protostellar evolution. 


In this paper, we aim to resolve this question by performing a systematic study 
of how PMS evolutionary tracks change as we alter the accretion history, 
the initial models, and the thermal efficiencies of mass accretion. 
We perform a systematic survey of parameter space in order to understand
how each of these factors affects protostellar evolution. This enables us
to answer the question of whether variation in any of 
these quantities could produce the appearance of an age spread in 
a population that is actually coeval.

The structure of the paper is as follows. 
In Section \ref{sec:method}, we briefly explain our 
numerical method for modeling protostellar evolution.
Section \ref{sec:results} is the main part of the paper, where
the numerical results are presented.
First, we investigate how different accretion histories influence
protostellar evolution in Section \ref{ssec:acchist}.
We next investigate protostellar evolution with differing
initial models in Section \ref{ssec:ini}, and with differing
thermal efficiencies in \ref{ssec:eff}. In Section \ref{sec:reliability}
we combine all these results to draw general conclusions about
the reliability of age and age spread estimates from PMS
evolutionary tracks.
Section \ref{sec:sum} contains the summary and discussion.

\section{Protostellar Evolution Models}
\label{sec:method}

We model protostellar evolution using the numerical code described 
by \citet{hosokawa09} and \citet{hosokawa10}. 
The code numerically solves the four stellar structure equations,
taking into account mass accretion. 
For the following calculations, we adopt the OPAL opacity tables
\citep{opal96} for high temperature $T > 7000$~K, and other tables 
based on the work by \citet{AF94} for the lower temperature.
We employ mixing-length theory for heat transport in
convective layers with a constant ratio of the
mixing length to the pressure scale height of 1.5.
We confirmed that our code reproduces the calculations
by \citet{sst80}, \citet{PS90}, and \citet{PS92} in both the limits of
hot spherical accretion and cold disk accretion, which we
explain in more detail below
(see appendixes in \citealt{hosokawa09,hosokawa10}).

We refer the reader to the 
Hosokawa et al.\ papers for full details of the numerical method, 
but one parameter is particularly important for the results
of this paper.
The thermal efficiency of mass accretion, i.e., the
entropy carried into the star with accreting material,
is a key parameter for protostellar evolution.
Since a protostar grows by accretion, the 
average entropy in the stellar interior becomes higher with
higher thermal efficiency.
For a star with a fixed mass, the stellar radius is larger for
higher interior entropy content.
Thus, we naively expect that, even for fixed accretion history,
protostars will have larger radii if the accretion flows onto them have
higher thermal efficiencies.

Despite efforts in previous work, however, the concrete value of 
the thermal efficiency in low-mass star formation is not well-constrained.
Here we address this uncertainty by considering two limiting cases,
representing cartoon versions of two
different accretion flow geometries: 
``hot" spherical accretion, and ``cold'' disk accretion.
In the hot accretion case, we envision that an accretion flow directly hits 
the stellar surface and forms an accretion shock front.
The accretion flow may arrive in a disk, but in the hot case we imagine
that the disk is thick enough so that
the accretion column covers much of the stellar surface.
As a result, a small fraction of the
heat generated at the shock front is carried into the stellar interior.
In this limit we solve for the steady structure of the gas accretion envelope 
as well as the stellar interior; the two are connected
at the stellar surface with accretion shock jump conditions 
\citep[e.g.,][]{sst80, hosokawa09}.
In contrast, in the cold accretion case we envision that accreting gas
initially falls onto a circumstellar disk and then reaches the stellar 
surface via a thin accretion column connecting the disk and star. 
As a result, most of the stellar 
photosphere is not covered by an accretion column and is able to radiate freely.
Accreting gas softly settles on the stellar surface, and when 
it is incorporated into the star it has the same entropy as gas  
in the stellar photosphere. 
In this case we do not solve for the structure of the accretion flow,
and we instead adopt the ordinary photospheric boundary condition  
(e.g., \citealt{PS92, hosokawa10}).

Our treatment of boundary conditions differs slightly from that of BCG09,
who modeled the thermal efficiency with a parameter
$\alpha$, the fraction of accreting internal energy 
absorbed by the star.
However, our limiting cases of hot and cold accretion 
just correspond to their $\alpha = 1$ and $0$ cases respectively.
The only other difference between our and BCG09's method is
that BCG09 assume instantaneous and uniform mixing of accreting 
material in the stellar interior \citep{siess96}. 
In this case, the entropy of newly accreted material is assumed to
be the same as the local values in the stellar interior, whereas
in our cold case it is assumed to match the stellar photosphere.
Note that, even in the hot case, only a small fraction of the
accretion energy goes into heating the stellar matter. 
This is similar to the $\alpha = 1$ case in BCG09, where 
the accretion energy acts as a uniform heating source
that is distributed uniformly throughout the stellar interior
\citep[e.g., see][]{siess97}.
As a result of this treatment,
most of this energy escapes from the star
without being absorbed by the stellar matter.
Thus, the term ``hot mass accretion,'' in both our treatment
and BCG09's, indicates that
mass accretion increases the average entropy in the stellar interior,
not that the accretion flow is completely radiatively inefficient.

\section{Results}
\label{sec:results}

\begin{deluxetable*}{cccccccc}
\tablecaption{
Model parameters and results
\label{tb:md}
}
\tabletypesize{\scriptsize}
\tablehead{
\colhead{Case} &
\colhead{Accretion History} &
\colhead{Boundary Condition} &
\colhead{$R_{*,0}$ ($R_\odot$)} &
\colhead{$M_{*,d}$ ($M_\odot$)} &
\colhead{$M_{*,f}$ ($M_\odot$)} &
\colhead{$R_{*,f}$ ($R_\odot$)} &
\colhead{$t_f$ (kyr)} 
}
\startdata
\cutinhead{1a. Fixed Initial and Boundary Conditions, Varying Accretion History}
mC5-C    & $10^{-5}~\msunyr$ & C & 1.5  & 0.074 & 0.9  & 1.3 
        & 90 \\                                               
mE-C     & episodic    & C   & 1.5  & 0.07 & 0.9 & 1.8 
        & 90 \\ 
mO-C     & simulation\tablenotemark{a}  & C   & 1.5  & 0.075 & 0.45 & 1.3
        & 110 \\
mOx2-C   & simulation\tablenotemark{a} & C   & 1.5  & 0.077 & 0.9  & 1.1
        & 110  \\
mOx0.5-C & simulation\tablenotemark{a} & C   & 1.5  & 0.073 & 0.23 & 1.3
        & 110 \\
mC4-C    & $10^{-4}~\msunyr$ & C & 1.5  & 0.076  & 0.9  & 0.92 
        & 9  \\                                               
mC6-C    & $10^{-6}~\msunyr$ & C & 1.5  & 0.1  & 0.9  & 1.4  
        & 900  \\                                               
\cutinhead{1b. Fixed Accretion History and Boundary Conditions,
Varying Initial Conditions}
mC5-C-Ri-8   &  $10^{-5}~\msunyr$  &  C & 8.0  & 0.093 & 0.9  
           & 1.8  & 90  \\
mC5-C-Ri3.7   &  $10^{-5}~\msunyr$  &  C & 3.7  & 0.09 & 0.9  
           & 1.7  & 90  \\
mC5-C       & $10^{-5}~\msunyr$ & C & 1.5  & 0.074 & 0.9  & 1.3 
           & 90 \\
mC5-C-Ri1   &  $10^{-5}~\msunyr$  & C & 1.0  & 0.06  & 0.9 & 1.1   
              & 90  \\                                                    
mC5-C-Ri0.65   &  $10^{-5}~\msunyr$  & C & 0.65  & 0.049  & 0.9 & 0.85    
              & 90  \\   
mC5-C-Ri0.3   &  $10^{-5}~\msunyr$  & C &  0.3  &  0.035 & 0.9 &  0.64  
           & 90  \\                                                    
mC5-C-Ri0.25   &  $10^{-5}~\msunyr$  & C & 0.25  & 0.033 & 0.9 & 0.4   
              & 90  \\                                                    
mC5-C-Ri0.2   &  $10^{-5}~\msunyr$  & C & 0.2  & 0.033 & 0.9 & 0.29   
              & 90  \\   
\cutinhead{1c. Fixed Accretion History and Initial Conditions, 
Varying Boundary Conditions}
mC5-C-Ri3.7   &  $10^{-5}~\msunyr$  &  C & 3.7  & 0.09 & 0.9  
           & 1.7  & 90  \\
mC5-C       & $10^{-5}~\msunyr$ & C & 1.5  & 0.074 & 0.9  & 1.3 
           & 90 \\
mC5-H       &  $10^{-5}~\msunyr$  & H  & 3.7  &  0.34 & 0.9  & 4.6
           & 90  \\
mC5-HC0.3   &  $10^{-5}~\msunyr$  &  H $\to$ C
($0.3~M_\odot$)\tablenotemark{b} & 3.7 (3.0)\tablenotemark{b}  & 0.33 & 0.9
           & 4.3  & 90  \\
mC5-HC0.1   &  $10^{-5}~\msunyr$  &  H $\to$ C ($0.1~M_\odot$)\tablenotemark{b} & 3.7 (2.6)\tablenotemark{b}  & 0.25 & 0.9 
           & 3.6  & 90  \\
mC5-HC0.03   &  $10^{-5}~\msunyr$  &  H $\to$ C
($0.03~M_\odot$)\tablenotemark{b} & 3.7 (3.2)\tablenotemark{b}  & 0.17 & 0.9 
           & 2.9  & 90  \\
\enddata
\tablecomments{
Col.\ 2: see main text for details of the accretion histories we use;
Col.\ 3: H = hot accretion, C = cold accretion; Col.\ 4: initial stellar 
radius when $M_*=0.01$ $\msun$; Col.\ 5: stellar mass when core
deuterium burning begins; Col.\ 6: final stellar mass for the most
 massive star we produce with these conditions; for most cases we also 
run to a series of smaller final masses; Col.\ 7: stellar radius at the 
end of accretion for the most massive case we run; Col.\ 8: time when 
accretion ends for the most massive case we run}
\tablenotetext{a}{For run mO-C the accretion history is taken from the 
simulations of \citet{offner09} (see text for details). Runs mOx0.5-C 
and mOx2-C use the same accretion history, scaled by factors of 0.5 and
 2, respectively.}
\tablenotetext{b}{For run mC5-HC$x$, the boundary condition is switched
from hot to cold once the stellar mass reaches $x~\msun$. 
The quantities given in parentheses are the stellar mass and radius 
when this switch occurs.}
\end{deluxetable*}

\subsection{Accretion History Variation}
\label{ssec:acchist}

\begin{figure}
 \begin{center}
\epsscale{1.0}
\plotone{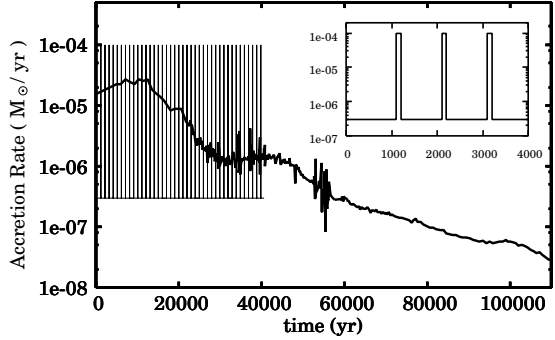}
\caption{ Time-dependent accretion histories adopted for 
calculations of protostellar evolution.
The thick solid line presents the sample accretion 
history taken from the numerical simulations of low-mass star 
formation by \citet{offner09}, and used in model mO-C.
The thin solid line shows the vigorous episodic accretion case
used in model mE-C, where a burst-like accretion phase at $10^{-4}~\msunyr$
over 100 years is interspersed with a quiescent phase at 
$3 \times 10^{-7}~\msunyr$ over 1000 years. The accretion history 
over initial 40000 years is shown for this case.
The small window enlarges the evolution over the initial 4000 years.}
\label{fig:acchist}
 \end{center}
\end{figure}

First, we examine how variations in protostellar accretion histories 
affect stars' HRD evolution.
To this end, we calculate a series of models, summarized in 
Table \ref{tb:md}a. 
In order to isolate dependence on the accretion history from other effects,
all these calculations use the same initial model and boundary conditions. 
This initial model consists of a $0.01~M_\odot$ star
with radiative interior, as in \citet{sst80}, and an initial radius 
of $1.5~R_\odot$.
Note that the value of $1.5~R_\odot$ is a bit smaller than the 
radius of the seed protostar calculated by \citet[hereafter MI00]{masunaga00}, 
$4~R_\odot$.
The boundary condition for all these models is cold accretion. 
We note that BCG09 obtained protostellar evolutionary tracks
that deviate from the non-accreting isochrones substantially only
for their cold cases (their $\alpha = 0$), and this motivates us to focus
on cold accretion first.

\begin{figure}
 \begin{center}
\epsscale{1.0}
\plotone{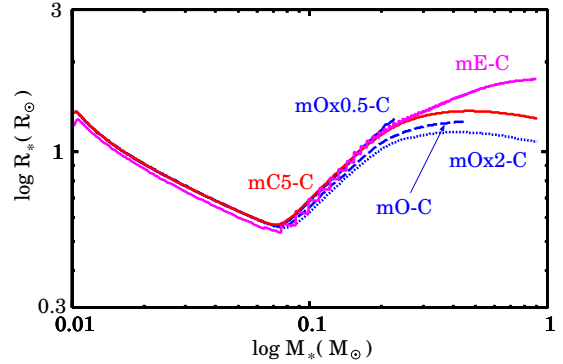}
\caption{ Evolution of the stellar radius versus stellar mass
for cases mC5-C (red), mE-C (magenta), mOx0.5-C (blue solid),
mO-C (blue dashed), and mOx2-C (blue dotted), among which the accretion
histories differ but the initial and boundary conditions do not. 
}
\label{fig:mr_acc}
 \end{center}
\end{figure}
\begin{figure*}
 \begin{center}
\epsscale{1.0}
\plotone{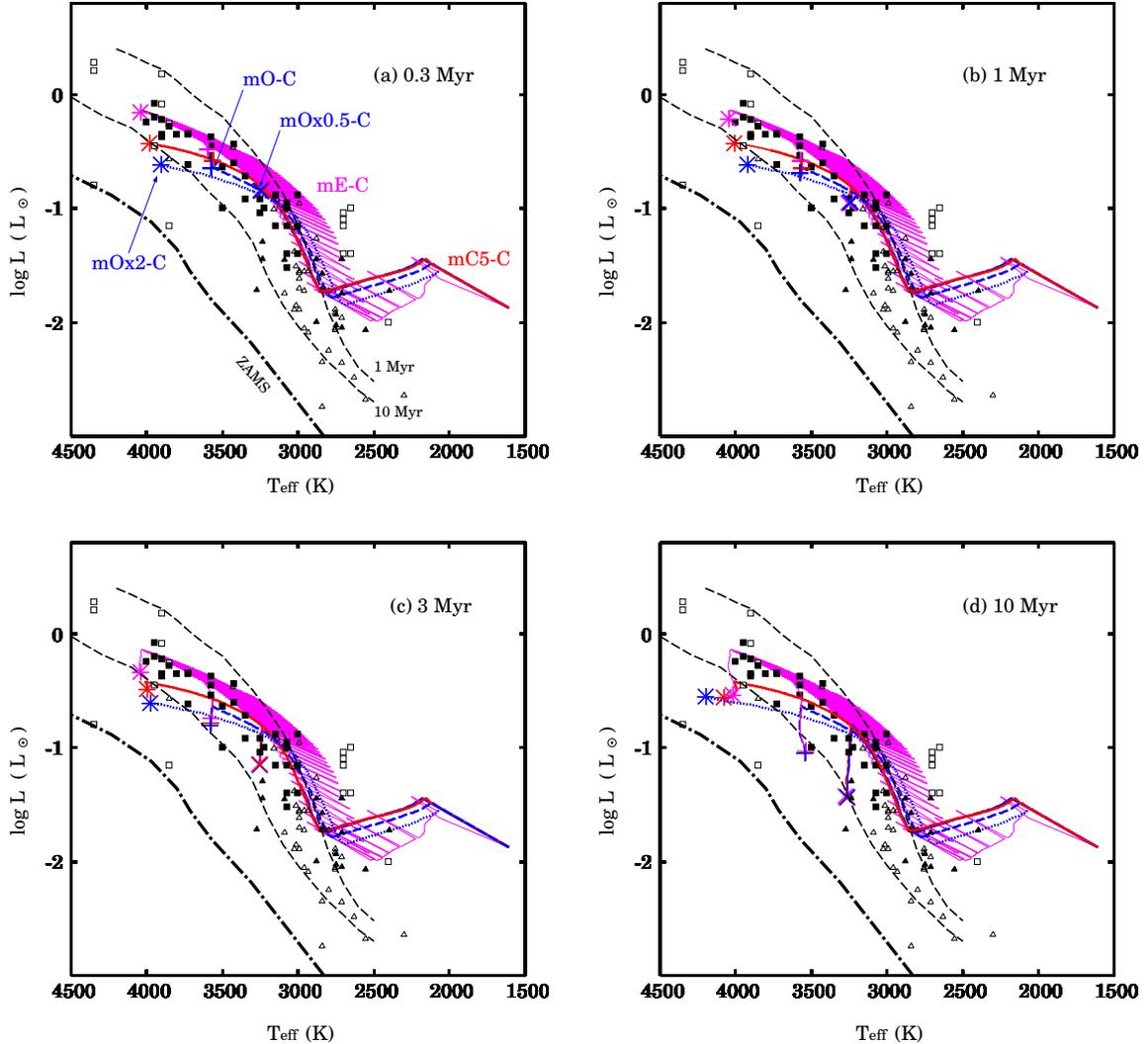}
\caption{
Snapshots of stellar positions in the HRD for varying accretion 
histories with the same initial and boundary conditions.
The four panels show the snapshots at times 
0.3~Myr (upper left), 1~Myr (upper right), 3~Myr (lower left),
and 10~Myr (lower right) after the start of mass accretion.  
The evolutionary tracks until that time are also plotted
in the panels.
(Note that the luminosity plotted here is only the stellar
luminosity; accretion luminosity is not included.)
The different colors represent different accretion histories:
constant accretion at $10^{-5}~\msunyr$ (red, case mC5-C),
episodic accretion (magenta, case mE-C), and decreasing accretion 
(blue, cases mO-C, mOx2-C, mOx0.5-C).
The input parameters in each case are summarized in Table \ref{tb:md}a.
In each panel, the symbols mark the positions of stars whose
masses are $0.9~M_\odot$ (asterisks), $0.45~M_\odot$ (pluses),
and $0.23~M_\odot$ (crosses). 
The thick dot-solid line indicates the positions of ZAMS stars 
\citep{siess00}. 
The dashed lines represent the isochrones of 1~Myr and 10~Myr 
for non-accreting protostars \citep{baraffe98}.
Observational data is taken from \citet[][open squares]{gatti06}, 
\citet[][filled squares]{gatti08}, 
\citet[][open triangles]{muzerolle05}, and 
\citet[][filled triangles]{peterson08}. 
}
\label{fig:HRacc}
 \end{center}
\end{figure*}
\begin{figure*}
 \begin{center}
\epsscale{1.0}
\plotone{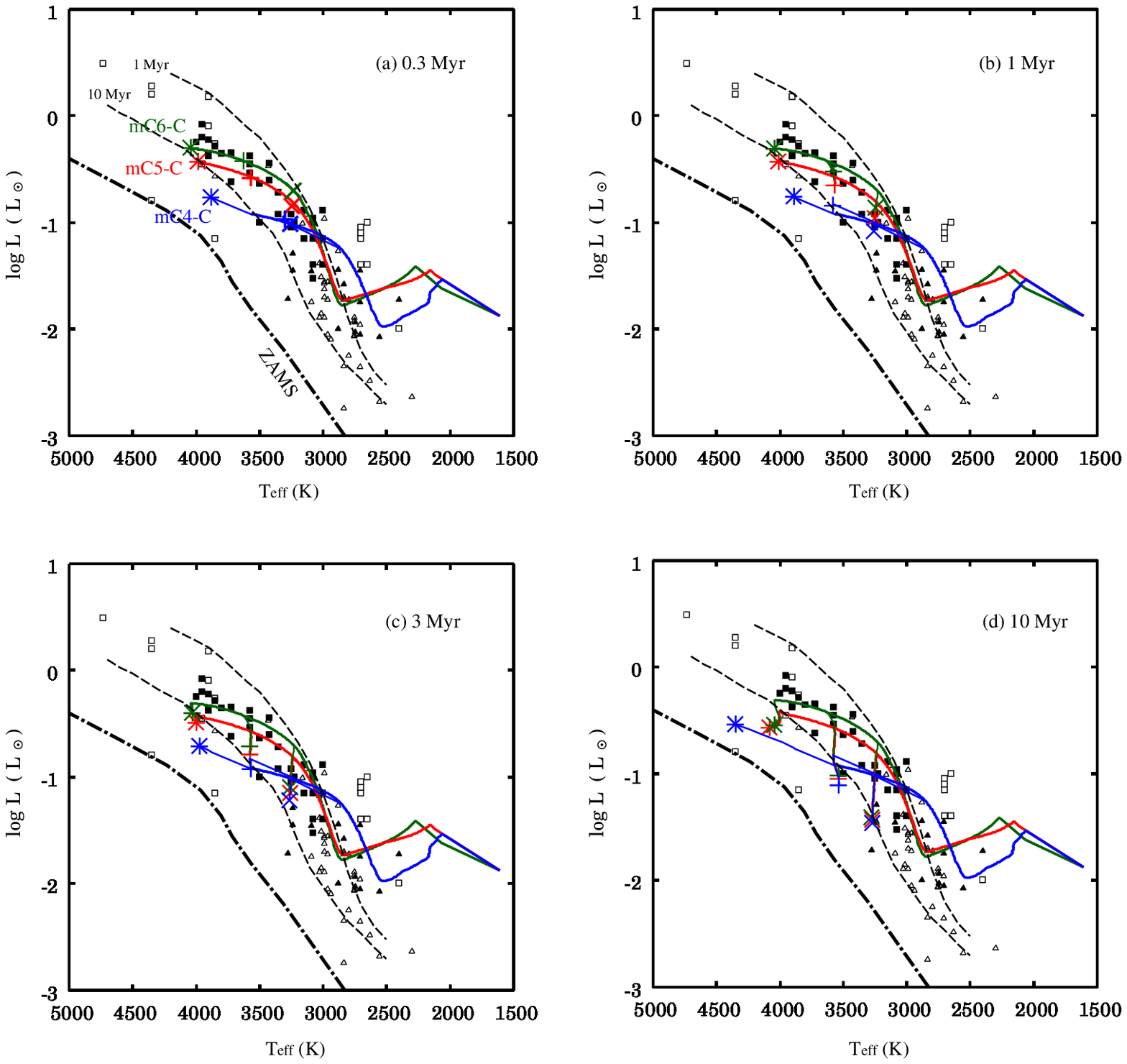}
\caption{Same as Figure \ref{fig:HRacc} but for different constant 
accretion rates. The different colors donate differences of the 
accretion rates: $10^{-6}~\msunyr$ (green, case mC6-C), 
$10^{-5}~\msunyr$ (red, case mC5-C), and 
$10^{-4}~\msunyr$ (blue, case mC4-C). Parameters in each case are
summarized in Table \ref{tb:md}a. }
\label{fig:HRxmdot}
 \end{center}
\end{figure*}

\subsubsection{Episodic Accretion}

We first consider five distinct accretion histories
with varying degrees of episodic variation, but with only factor
of $\sim 2$ level changes in the average accretion rate. This enables
us to explore the sensitivity of stars' HRD evolution to the level of
variability in their accretion histories.
Model mO-C uses an
accretion history taken from numerical simulations of low-mass star 
formation by \citet{offner09} and is illustrated in Figure
\ref{fig:acchist}. In this case, the accretion rate gradually decreases over 
$\simeq 0.1$~Myr, and the stellar mass finally reaches 0.45~$M_\odot$. 
Models mOx0.5-C and mOx2-C use the same accretion history, but they 
are scaled by factors of 0.5 and 2, respectively, to give final masses 
of $0.23~M_\odot$ and $0.9~M_\odot$. In contrast, model mC5-C uses a 
fixed accretion rate of $\dot{M} = 10^{-5}~\msunyr$. 
We use this model to produce $0.23$, $0.45$, and $0.9$ $\msun$ stars, 
as in the mO-C models, simply by stopping accretion once
the star has reached the desired mass. Finally, model mE-C (also shown 
in Figure \ref{fig:acchist}) represents an extreme case of variability: 
an episodic mass accretion history
where burst-like accretion events at $\dot{M} = 10^{-4}~\msunyr$ 
over 100 years are interspersed with quiescent phases of accretion at 
$3 \times 10^{-7}~\msunyr$ over 1000 years. This model is similar to
the episodic accretion histories formulated by BCG09.
As with mC5-C, we use
this model to produce $0.23~M_\odot$, $0.45~M_\odot$, and $0.9~M_\odot$
stars simply by turning off mass accretion once the stellar mass reaches
the target value.

Figure \ref{fig:mr_acc} presents the evolution of the stellar radius
until mass accretion ceases in each case.
We see that the basic evolution is similar for all cases.
The stellar radius initially decreases with increasing 
stellar mass. The temperature in the stellar interior
rises during this initial contraction.
When the stellar mass reaches $M_* \simeq 0.07~M_\odot$, deuterium
burning begins and the stellar interior becomes fully convective.
For some time after the ignition of deuterium, temperature at the
stellar center remains constant at $\simeq 10^6$~K due to
the very strong temperature-dependence of the deuterium burning rate.
This is the so-called thermostat effect of deuterium burning 
\citep[e.g.,][]{stahler88}.
The stellar radius increases in proportion to the stellar mass
during this phase.
The deuterium concentration in the stellar interior significantly
decreases with increasing the stellar mass \citep{stahler88, hartmann97}.
Finally at $M_* \gtrsim 0.2~M_\odot$, the deuterium concentration
is so low that the thermostat effect becomes inoperative and the 
central temperature increases again.
Variation of the accretion histories only slightly 
influences the evolution of the stellar radius.

\setcounter{footnote}{0}

Figure \ref{fig:HRacc} 
\footnote{
Note that this figure is intended to facilitate comparison between the models
and data for PMS stars, for which
mass accretion has presumably ceased. 
In this and all subsequent figures we omit the accretion
luminosity for the early
evolutionary tracks.
If the accretion luminosity is added, the tracks shift upward 
in the HRD and no longer pass through the data.}
shows the stellar positions in the HRD at 
$t =0.3$~Myr, 1~Myr, 3~Myr, and 10~Myr after mass 
accretion begins for each mass accretion history.
The snapshot at $t = 0.3$~Myr shows the stellar positions just 
after mass accretion ceases.
Reflecting the minimal variation in the stars' radii
shown in Figure \ref{fig:mr_acc}, the stellar positions only show 
a small spread.
This means that the concept of the birthline is valid even with 
variable accretion histories in the limiting case of cold mass
accretion, provided the initial state is the same from star to star.
At $t > 0.3$~Myr, the stars gradually approach the ZAMS line, 
descending in the HRD 
(\citealt{henyey55, hayashi61, hayashi_nakano63}). 
The snapshot at $t = 1$~Myr clearly shows that the stars are
below the 1~Myr isochrone for non-accreting protostars.
This offset is larger for the higher-mass stars.
In particular, the $0.9~M_\odot$ stars are close to the 
isochrone of 10~Myr.
The offset decreases with time, but still remains even
at $t = 10$~Myr for $0.9~M_\odot$ stars.

This divergence between the accreting evolutionary histories
and the non-accreting isochrones is easy to understand.
The isochrones for non-accreting protostars are derived assuming
a large initial radius of $R_{*,0} \sim 10~R_\odot$.
The model PMS stars then contract from this initial state by radiating 
away their energy, reaching smaller radii at larger stellar ages.
On the other hand, with thermally inefficient accretion the 
stellar radius remains as small as $R_* \lesssim 1~R_\odot$ during
mass accretion.
As a result, stellar ages are overestimated using the 
isochrones for non-accreting protostars in such cases.
However, we stress that this effect has nothing to do with
the time variability of the mass accretion rate.

\subsubsection{Varying Mean Accretion Rates}

Next we examine protostellar evolution over a 
greater range of mass accretion rates but with no
time variability. In addition to case mC5-C, we 
consider cases mC4-C and mC6-C, which have constant 
accretion rates of $10^{-4}~\msunyr$ and $10^{-6}~\msunyr$, respectively.
Figure \ref{fig:HRxmdot} is the same as Figure \ref{fig:HRacc} but for
these cases. 
We see larger variation of the tracks among these
cases than in Figure \ref{fig:HRacc}, but the level of
variation is still significantly smaller than the observed
range of data, particularly for the lowest mass and effective
temperature. There is little difference between cases mC5-C and mC6-C,
where accretion rates take typical values for
low-mass star formation of $\dot{M} \lesssim 10^{-5}~\msunyr$.
The mC4-C track slightly deviates from these tracks.
In this case, however, the $0.9~M_\odot$ star always lies below the 10~Myr 
isochrone, far from the locations where observed protostars lie.
We therefore conclude that purely cold accretion at
rates of $10^{-4}~\msunyr$, while not physically forbidden,
does not appear to actually occur in observed star clusters, at least for
$0.9~M_\odot$ stars.
The problem that many evolutionary scenarios involving cold
accretion overpopulate the region at low $L$ and high $T_{\rm eff}$
is one we will encounter repeatedly in the rest of this paper.

In summary, we conclude that 
star-to-star differences in either the overall accretion
rate or the degree to which the accretion rate varies in time have
a very limited effect on protostellar evolution.
As illustrated by Figures \ref{fig:HRacc} and \ref{fig:HRxmdot},
varying accretion histories can cause absolute age estimates to be wrong for 
more massive stars, but {\it cannot} explain the observed broad spread 
of PMS stars in the HRD. 
If PMS stars' initial state and accretion 
boundary conditions were fixed, then a co-eval population would
form a much tighter sequence in the HRD than what we
actually observe, even if their accretion
histories varied wildly.

\subsection{Initial Model Variation}
\label{ssec:ini}

\begin{figure}
 \begin{center}
\epsscale{1.0}
\plotone{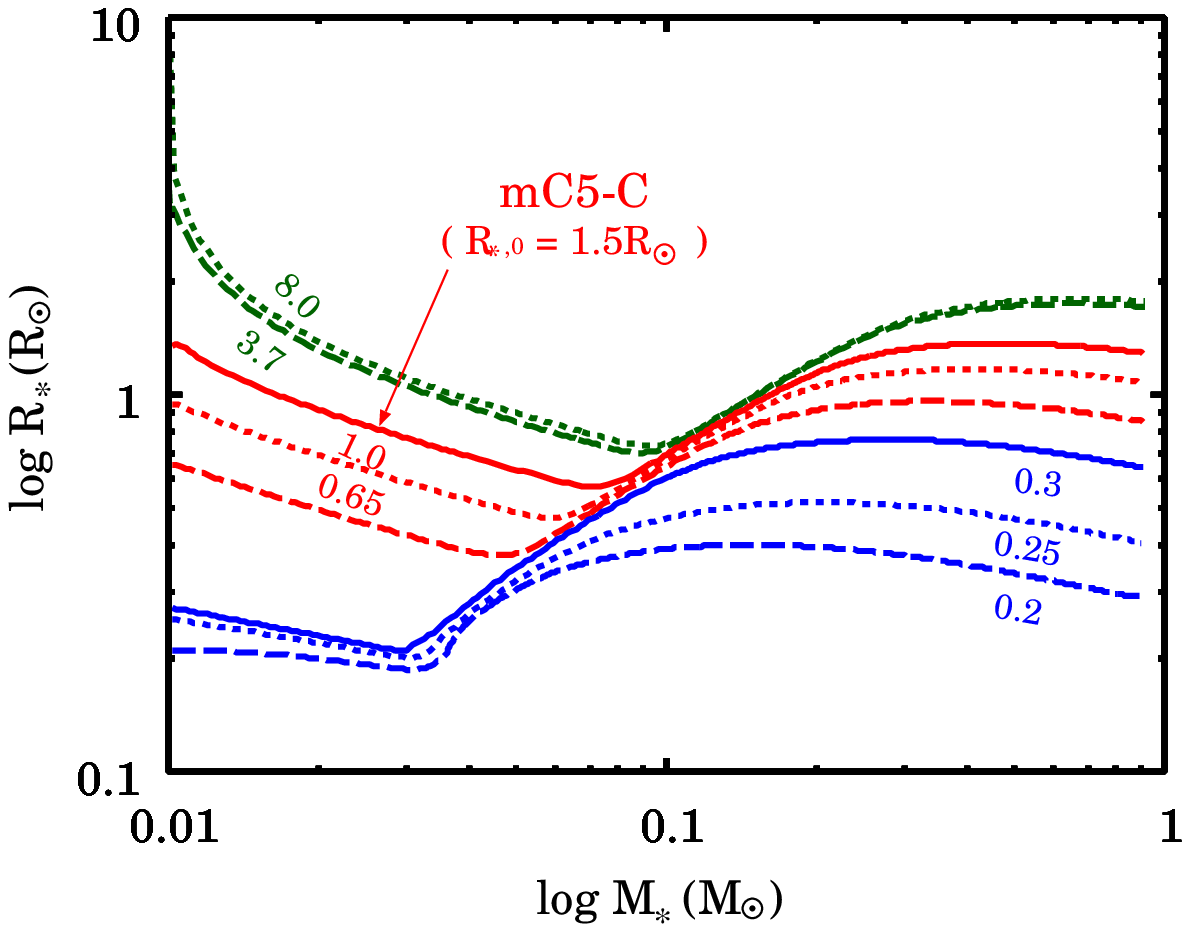}
\caption{ 
Same as Fig. \ref{fig:mr_acc} but for cases with different initial
radii: $R_{*,0} = 8.0~R_\odot$ (green dotted, case mC5-C-Ri8),
$R_{*,0} = 3.7~R_\odot$ (green dashed, case mC5-C-Ri3.7),
$R_{*,0} = 1.5~R_\odot$ (red solid, case mC5-C), $1.0~R_\odot$
(red dotted, case mC5-C-Ri1), $0.65~R_\odot$ (red dashed, case mC5-C-Ri0.65), 
$0.3~R_\odot$ (blue solid, case mC5-C-Ri0.3), $0.25~R_\odot$ 
(blue dotted, case mC5-C-Ri0.25), 
and $0.2~R_\odot$ (blue dashed, case mC5-C-Ri0.2).
The values of the initial radii are labeled in the panel. 
The cold mass accretion at the constant rate 
$\dot{M} = 10^{-5}~\msunyr$ is adopted for all the cases.}
\label{fig:mr_ini}
 \end{center}
\end{figure}
\begin{figure}
 \begin{center}
\epsscale{1.0}
\plotone{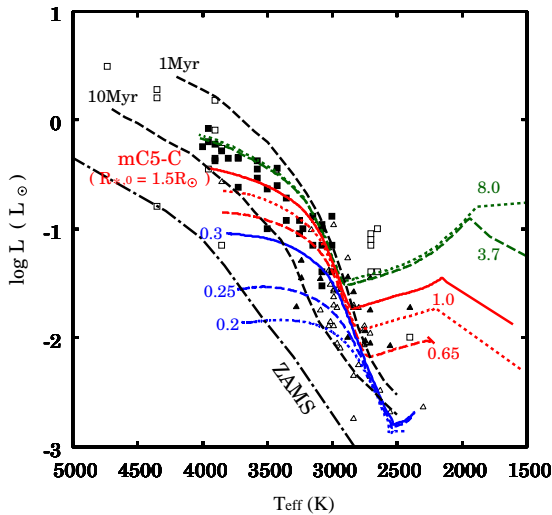}
\caption{
The evolution tracks with varying the initial radius with
the cold mass accretion. 
The constant accretion rate of $10^{-5}~\msunyr$ is adopted 
for these all cases. 
The tracks until the protostellar mass
reaches $0.9~M_\odot$ by mass accretion are presented.             
The different lines shows the evolution with different initial
radii: 
$R_{*,0} = 8.0~R_\odot$ (green dotted, case mC5-C-Ri8),
$R_{*,0} = 3.7~R_\odot$ (green dashed, case mC5-C-Ri3.7),
$R_{*,0} = 1.5~R_\odot$ (red solid, case mC5-C), 
$1.0~R_\odot$ (red dotted, case mC5-C-Ri1), 
$0.65~R_\odot$ (red dashed, case mC5-C-Ri0.65), 
$0.3~R_\odot$ (blue solid, case mC5-C-Ri0.3), 
$0.25~R_\odot$ (blue dotted, case mC5-C-Ri0.25), and
$0.2~R_\odot$ (blue dashed, case mC5-C-Ri0.2).
The initial radius in each model is labeled in the figure.
The input parameters for these models are summarized in Table \ref{tb:md}b.}
\label{fig:ri_track}
 \end{center}
\end{figure}
\begin{figure*}
 \begin{center}
\epsscale{1.0}
\plotone{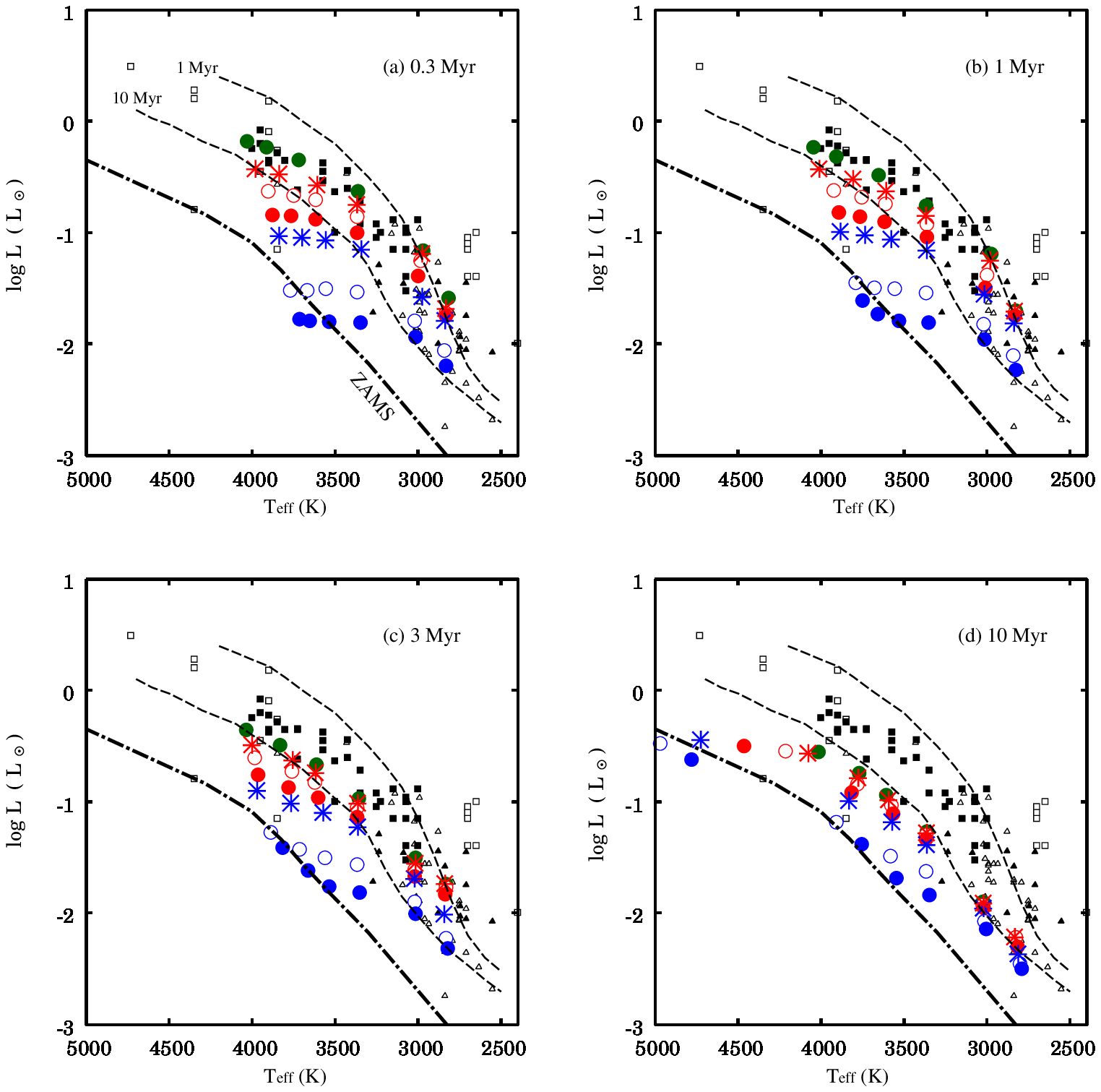}
\caption{
Same as Fig. \ref{fig:HRacc} but for the cases presented in
Fig. \ref{fig:ri_track} and for the PMS stars of various masses. 
The symbols in each panel mark the positions of $0.9~M_\odot$,
$0.7~M_\odot$, $0.5~M_\odot$, $0.3~M_\odot$, $0.1~M_\odot$,
and $0.05~M_\odot$ stars from the left for cases 
mC5-C-Ri3.7 (green filled circles),
mC5-C (red asterisks), 
mC5-C-Ri1 (red open circles), mC5-C-Ri0.65 (red filled circles), 
mC5-C-Ri0.3 (blue asterisks), mC5-C-Ri0.25 (blue open circles), 
and mC5-C-Ri0.2 (blue filled circles).
We omit the evolutionary tracks for clarity here. We also omit
model mC5-C-Ri8 because it is nearly identical to mC5-C-Ri3.7.}
\label{fig:ri_pms}
 \end{center}
\end{figure*}

We next study protostellar evolution by varying the initial
model, while fixing the boundary condition to cold mass accretion
and the accretion rate to $\dot{M} = 10^{-5}~\msunyr$.
We choose cold accretion here, because
protostellar evolution is only sensitive to the initial model in 
the cold case, not the hot one \citep{stahler88, hartmann97} --
we defer a detailed discussion of this issue to Section \ref{ssec:eff}.
Throughout these calculations, we assume a fixed initial
deuterium abundance, which in principle may vary somewhat. 
\citet{stahler88} explored the effect of varying deuterium abundance 
on the stellar radius. He showed that larger amounts of deuterium 
led to somewhat larger stars. However, this had only a small 
effect on the total thermal energy due to the thermostat effect 
of deuterium burning, which acts to regulate the
central stellar temperature.

Unlike the accretion history, which is a macroscopic property
that almost certainly varies from star to star, the initial radius
is at least partly fixed by microphysics. The initial model should
correspond to the ``seed'' protostar which forms as a result 
of second collapse induced by collisional dissociation of
hydrogen molecules in a thermally-supported first core
(e.g.\ \citealt{Larson69, WN80, MMI98}; MI00).
The entropy content of the resulting seed protostar, and
thus its initial radius, is 
therefore at least partly set by the properties of the
hydrogen molecule, in which case we would not expect
large star-to-star variations.
Nonetheless, models of second collapse have not extensively
explored the influence of factors like rotation or magnetic
fields. 
Thus, to be conservative we
consider a factor of $\sim 100$ variation in possible initial
radii, extending to values both larger and smaller
than the radius of $4~R_\odot$ computed by MI00. The
smallest radii we consider are much smaller than have
been produced in any calculation of second collapse. We list
the full set of models we examine in Table \ref{tb:md}b.

Figure \ref{fig:mr_ini} presents the evolution of the stellar radius
until mass accretion ceases in each case. 
We see that for all stellar masses these cases vary much more than 
those in Figure \ref{fig:mr_acc}.
With the smaller initial radius, the stellar radius is also smaller
after the stellar mass increases by mass accretion.
In case mC5-C-Ri0.3, for example, the temperature in the stellar 
interior is initially higher compared to case mC5-C.
As a result, deuterium burning begins earlier at $M_* \simeq 0.03~M_\odot$, 
and the stellar radius is always smaller than $1~R_\odot$.

Figure \ref{fig:ri_pms} shows the evolutionary tracks in the HRD 
until the stellar mass reaches $0.9~M_\odot$.
We see that tracks with small initial radii occupy the
lower part of the HRD, reflecting the variation shown
in Figure \ref{fig:mr_ini}.
Figure \ref{fig:ri_pms} shows the evolution of the stellar positions 
in the HRD after mass accretion ceases for these cases, computed
for stars that stop accreting at final masses of $0.05~M_\odot$, $0.1~M_\odot$, 
$0.3~M_\odot$, $0.5~M_\odot$, $0.7~M_\odot$, and $0.9~M_\odot$.
In comparison to Figures \ref{fig:HRacc} and \ref{fig:HRxmdot}, we see a 
much larger spread between the isochrones than that produced by
different accretion histories.
The stellar positions significantly differ even for the same mass and age.
We see that the observational data points near the 10~Myr non-accreting 
isochrone are covered even in the snapshots for $t \leq 1$~Myr. 

However, with the exception of models mC5-C-Ri3.7 and mC5-C-Ri8, 
which start with large initial radii and entropies, the distribution of 
the calculated PMS stars is never consistent with that of the
observational data points.
Even at the earliest time snapshot, in these models most or all the stars
with $T_{\rm eff} > 3500$ K lie below the 10 Myr isochrone, where
there are no observed stars.
This problem is particularly serious for the cases represented with 
blue symbols (cases mC5-C-Ri0.3, mC5-C-Ri0.25, and mC5-C-Ri0.2).

In order to render these cold accretion models with small radii consistent
with observations, one would have to posit that only stars
whose {\it final} masses are below $0.5~M_\odot$ have second cores
with radii much smaller than the values predicted by MI00 and similar
calculations. In effect, the $\sim 0.01~M_\odot$ second core would need
to know in advance 
what the final properties of the star would be, and the
properties of the second core would have to somehow correlate with the
final mass. 
Given the limited range of second collapse models that have been
explored in the literature, we cannot rule out the possibility that both
very small second core radii and a systematic correlation between
second core radii and final stellar masses exist in nature. However, we are
also unaware of any observational or theoretical evidence in favor
of either of these propositions.

If, on the other hand, we restrict our attention to models with initial 
radii such that the stars are at least marginally consistent with the 
data at all effective temperatures
(models mC5-CRi-8 to mC5-CRi0.65, shown in green and red in Figure 
\ref{fig:ri_pms}, we see that the spread in HRD location at 
$T_{\rm eff} < 3500$ K is very small, comparable to the spread seen 
in Fiugres \ref{fig:HRacc} and \ref{fig:HRxmdot}, and much smaller 
than the spread in observed stellar positions.

\subsection{Thermal Efficiency Variation}
\label{ssec:eff}

\begin{figure}
 \begin{center}
\epsscale{1.0}
\plotone{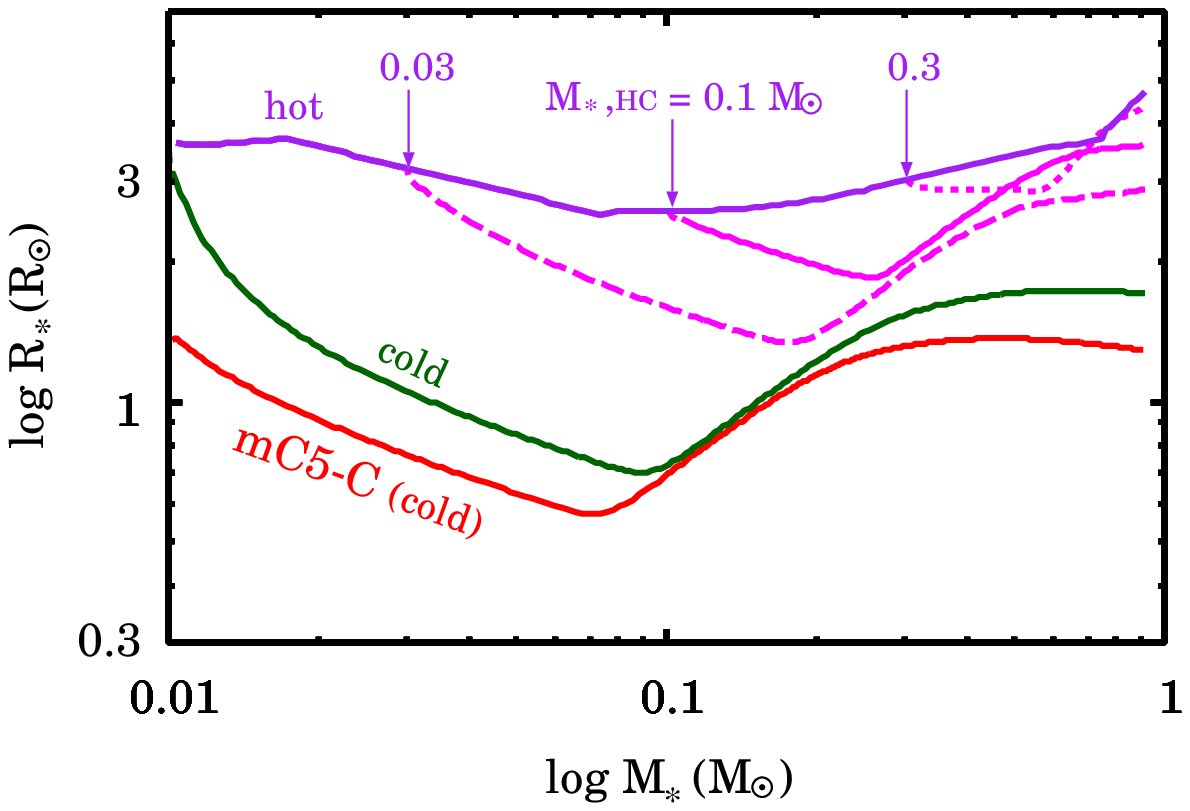}
\caption{Same as Fig. \ref{fig:mr_acc} but for cases with different
thermal efficiencies of mass accretion. 
Only hot and cold mass accretion is adopted with the same initial 
radius $3.7~R_\odot$ in cases mC5-H (purple) and mC5-C-Ri3.7 (green).
In the cases with the magenta lines, the hot mass accretion is 
switched to the cold accretion when the stellar mass reaches 
$M_{*,{\rm HC}} = 0.3~M_\odot$ (magenta dotted, case mC5-HC0.3), 
$0.1~M_\odot$ (magenta solid, case mC5-HC0.1), and 
$0.03~M_\odot$ (magenta dashed, case mC5-HC0.03).
The values of $M_{*,{\rm HC}}$ are labeled in the figure.  
The input parameters for these models are summarized in Table \ref{tb:md}c.
The red line represents the evolution in case mC5-C for a comparison.     
}
\label{fig:mr_h2c}
 \end{center}
\end{figure}
\begin{figure}
 \begin{center}
\epsscale{1.0}
\plotone{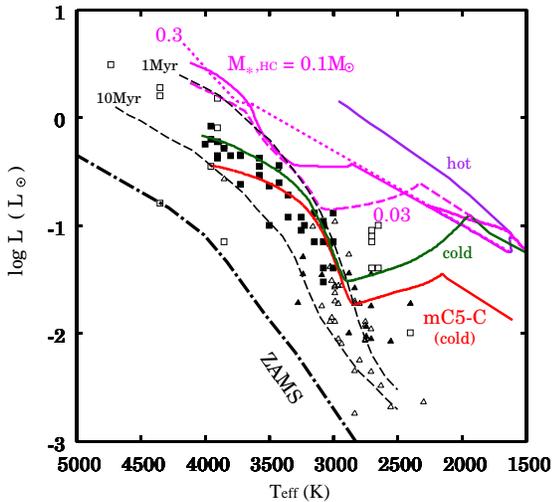}
\caption{Same as Fig. \ref{fig:ri_track} but for varying 
the thermal efficiency of mass accretion. 
Different lines indicate the different thermal efficiencies:
only hot mass accretion (purple, case mC5-H), cold accretion followed
by early hot accretion for $M_* < M_{*,{\rm HC}} = 0.3~M_\odot$
(magenta dotted, case mC5-HC0.3), 
$0.1~M_\odot$ (magenta solid, case mC5-HC0.1),  
$0.03~M_\odot$ (magenta dashed, case mC5-HC0.03),
and only cold mass accretion with the same initial model
as in the above cases (green, case mC5-C-Ri3.7)
}
\label{fig:h2c_track}
 \end{center}
\end{figure}
\begin{figure*}
 \begin{center}
\epsscale{1.0}
\plotone{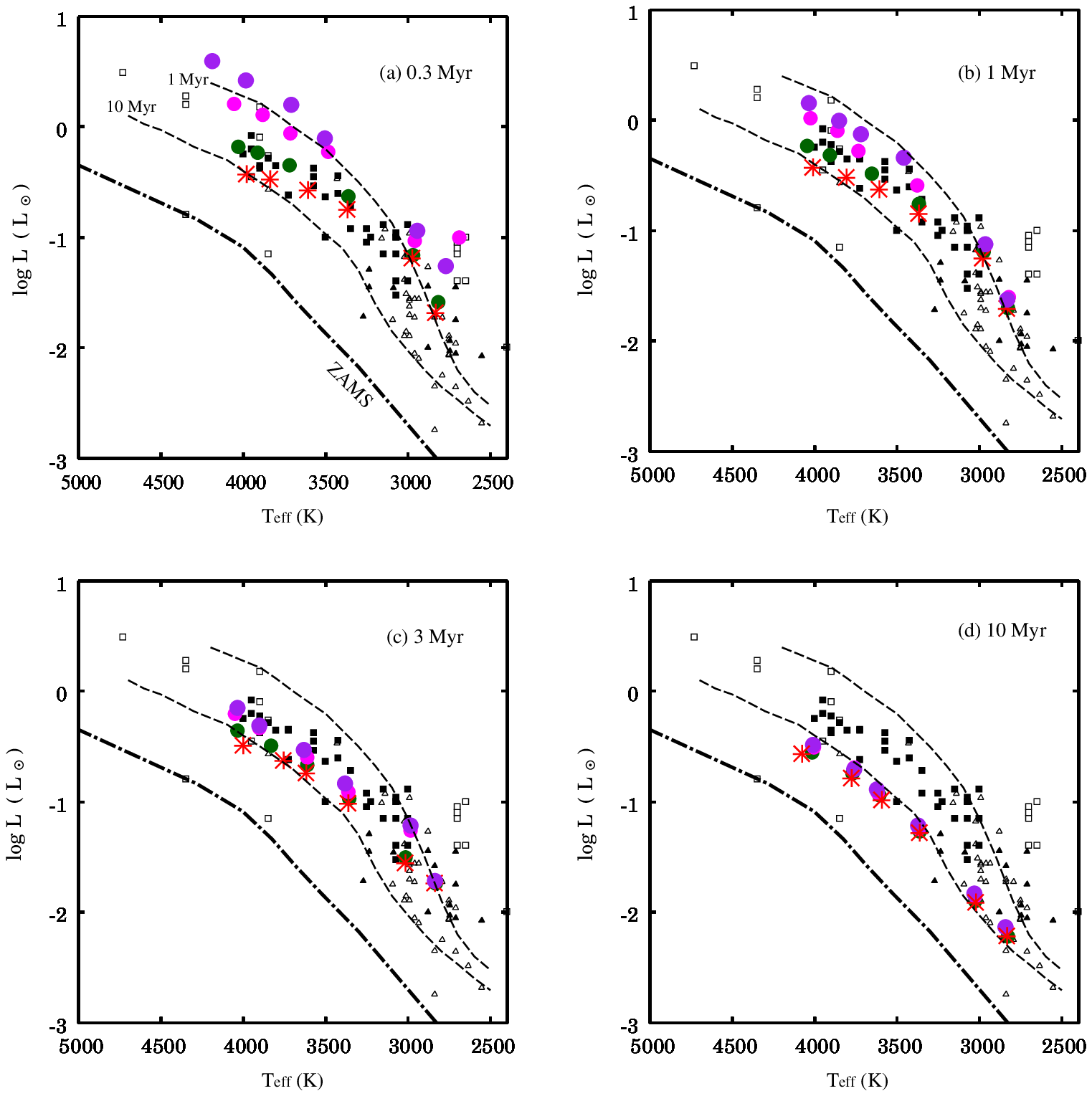}
\caption{Same as Fig. \ref{fig:ri_pms} but for the cases presented
in Fig. \ref{fig:h2c_track}.
The symbols in each panel mark the positions of $0.9~M_\odot$,
$0.7~M_\odot$, $0.5~M_\odot$, $0.3~M_\odot$, $0.1~M_\odot$,
and $0.05~M_\odot$ stars from the left for cases mC5-H 
(purple filled circles), mC5-HC0.03 (magenta filled circles),
mC5-C-Ri3.7 (green filled circles),  
and mC5-C (red asterisks).}
\label{fig:h2c_pms}
 \end{center}
\end{figure*}

Next, we consider the effects of different thermal efficiencies of
mass accretion on protostellar evolution.
In addition to the cold accretion models we have previously
considered we now add models that experience hot accretion 
for varying lengths of time during their evolution. We summarize
these models in Table \ref{tb:md}c. Model mC5-H uses pure
hot accretion, and models mC5-HC$x$ switch from hot to
cold accretion once their masses exceed $x~M_{\odot}$.

For these new hot models we adopt $0.01~M_\odot$ radiative stars
for the initial state, as in the previous section. We use initial
radii of $3.7~R_\odot$, which is roughly consistent with 
the value of $3.45~R_\odot$ in \citet{sst80} and the
value of $4.0~R_\odot$ computed by MI00, and somewhat
larger than the value of $2~R_\odot$ adopted in 
\citet{stahler88}.\footnote{The difference between
\citet{sst80} and \citet{stahler88} arises 
because \citet{stahler88} use fully convective initial models,
although the interior of a seed protostar would 
be radiative prior to deuterium burning \citep{sst80}. 
In our calculations, we follow the early evolution, where a 
convective layer occurs after the ignition of deuterium burning.}
However, the initial radius does not matter in the hot
case as it does in the cold one \citep{stahler88, hartmann97}.
In the cold case, the entropy of gas accreting onto the star
matches the entropy of the stellar atmosphere. Since the structure of
the stellar atmosphere depends on the 
initial model, so does the subsequent evolution. 
In contrast, for hot models there is a self-regulation mechanism
that removes the dependence on the initial radius.
If the initial radius is too small, accreting gas releases a large
amount of gravitational energy before reaching the stellar surface.
Since a fraction of this energy is trapped in the accreting gas 
in the hot case, accreting materials settle onto the star with high 
entropy, which increases the stellar radius. 
On the other hand, if the initial radius is too large,
the opposite effect operates. Accreting material has less
entropy because it converts less of its gravitational energy to
kinetic energy, and this serves to decrease the stellar radius.
The stellar radius is regulated as a result.

Figure \ref{fig:mr_h2c} shows the evolution of the stellar radius for 
the runs with varying thermal efficiency.
In the cases with the early hot accretion, the stellar radius at
$M_* = M_{*,{\rm HC}}$ is $\simeq 3~R_\odot$, significantly larger 
than in any of the purely cold cases considered in Section \ref{ssec:ini}.
In case mC5-HC0.1, for example, the stellar radius slightly 
decreases after the boundary condition is switched to cold 
mass accretion at $M_* = 0.1~M_\odot$. 
However, deuterium burning begins soon after and the star expands 
for $M_* \gtrsim 0.3~M_\odot$. 
The evolution at $M_* > M_{*,{\rm HC}}$ is close to that in the 
purely hot case mC5-H. Even with our lowest switching mass 
$M_{*,{\rm HC}} = 0.03~M_\odot$ (case MC5-HC0.03), the stellar
radius always exceeds $1~R_\odot$.

Figure \ref{fig:h2c_track} is the same as Figure \ref{fig:ri_track} 
but for the cases with varying efficiency.
We see that at the end of accretion all the models with any
hot accretion, even mC5-HC0.03
where we switch to cold accretion at $0.03$ $\msun$, lie at
or above the 1 Myr non-accreting isochrone, and above the
level of the observed data. The cold models are systematically lower.
Significantly, only the hot models are consistent with the stars
at the highest observed $L$ and $T_{\rm eff}$. We therefore conclude
that either these stars must experience some hot accretion, or that
their initial radii must be very large, thereby achieving the same effect.

Figure \ref{fig:h2c_pms} shows the subsequent time evolution of the stellar 
positions for these cases as in Figure \ref{fig:ri_pms}.
We omit cases mC5-HC0.1 and mC5-HC0.3 here, because they are
almost the same as case mC5-H. 
The snapshot at $t = 0.3$~Myr shows that, in case mC5-H, the stars 
are above the 1~Myr isochrone when mass accretion ceases.
The PMS stars descend in the HRD after this, and their positions
are nearly consistent with the isochrones at $t \gtrsim 1$~Myr.
The differences in stellar positions between cases mC5-H and mC5-HC0.03
are small even in the early snapshots when $t \leq 1$~Myr, 
and at $T_{\rm eff}\la 3500$ K the differences are particularly small.
Thus, even a short duration of hot mass accretion produces a 
similar result to the purely hot case.
BCG09 also find very bimodal outcomes between their hot cases
($\alpha \geq 0.2$) and cold cases ($\alpha = 0$).
This suggests that it is difficult to freely populate the PMS stars 
in the HRD only by varying the accretion thermal efficiency.

\section{The Reliability of Non-Accreting Isochrone Estimates
of Ages and Age Spreads}
\label{sec:reliability}

Having explored the parameter space thoroughly, we are
now in a position to make some general statements about the
reliability of theoretical isochrones as tools for estimating
ages and age spreads. In this analysis, it is helpful to 
separate the cases of stars with $T_{\rm eff} \ga 3500$ K
from those with $T_{\rm eff} \la 3500$, because the results
are quite different in the two cases.

At $T_{\rm eff} > 3500$~K, stars with high and low thermal 
efficiencies, or with different initial radii, 
can end up quite far apart in the HRD even at equal ages
(Figures \ref{fig:ri_pms} and \ref{fig:h2c_pms}). 
We therefore conclude that, in this regime, stellar age and mass alone 
cannot uniquely determine stellar positions in the HRD, and stellar 
age estimates based on HRD positions cannot in general be considered reliable. 
However, we note that errors in this regime only occur in one direction:
young stars can appear old, but old stars never appear young. In no case
do we find stars above the 1 Myr isochrone whose ages are actually
$>1$ Myr.
Thus very young age estimates are reliable in this effective temperature 
range, even if old age estimates are not. 
With regard to age spreads, we note that the snapshots at
$t \lesssim 1$~Myr show that the models in which we
vary the thermal efficiency and the initial radius cover the
entire observed spread of PMS stars with $T_{\rm eff} > 3500$~K.
Thus, the entire observed luminosity spread in this
temperature range could be explained if a coeval population
were to consist of some stars that underwent hot accretion
and others that underwent cold accretion. 
Estimates of age spreads in this regime are therefore unreliable.

The situation is quite different for $T_{\rm eff} \la 3500$ K.
In this regime, consulting Figures \ref{fig:HRacc}, \ref{fig:HRxmdot},
\ref{fig:ri_pms}, and \ref{fig:h2c_pms} shows that the only models
capable of producing false old ages 
are those that have pure cold accretion starting from very small
initial radii. For example, even at a true age of 3 Myr, only model
mC5-C-Ri0.2 places stars with $T_{\rm eff} = 3000$ K below the
10 Myr isochrone. However, these models can be considered acceptable 
and consistent with the observational data only if two uncertain propositions
hold. First, a significant number of stars would have to accrete
from seeds whose radii are an order of magnitude smaller than
any that have been produced in theoretical models thus far. Second,
in order to avoid drastically overpopulating  the region 
below the 10 Myr isochrone at $T_{\rm eff} \ga 3500$ K, 
such small initial radii would have to be realized only for stars
that have small final masses at the end of accretion. There would
have to be some mechanism to generate a correlation between
the radii of $\sim 0.01$ $\msun$ second cores and the masses
of the $\sim 1$ $\msun$ stars to which they eventually grow.
Given the limitations of the theoretical models, neither of these
propositions can be ruled out, but there is currently no evidence
in favor of them either.

If we rule out the cold accretion, small initial radius models
based on these problems, we find that
the observed luminosity spread for $T_{\rm eff} < 3500$~K is
considerably larger than the spread among the remaining models.
All these models lie fairly close to the non-accreting 
isochrones appropriate to their true ages for times
$t\gtrsim 3$ Myr, and at times $t\gtrsim 1$ Myr at $T_{\rm eff} \sim 3000$~K.
We therefore conclude that, unless $\sim 0.01$ $\msun$ seed 
protostars have some very specific unexpected properties,
age estimates of $1-3$ Myr or more based on
non-accreting isochrones are reliable for stars with effective
temperatures below $\sim 3500$~K, 
at least insofar as the observations themselves are reliable.
Neither variation in thermal efficiency, accretion history, or initial radius
can explain the observed spread in the HRD.

\section{Summary and Discussion}
\label{sec:sum}

In this paper, we have examined a variety of low-mass 
protostellar evolutionary tracks with varying accretion histories,
initial models, and thermal efficiencies of mass accretion. 
We have also compared the resultant
spread of PMS stars in the HRD to that observed in nearby
low-mass star forming regions 
\citep{peterson08, gatti06,gatti08, muzerolle05}. 

We first calculate protostellar evolution models with
varying accretion histories but fixed initial stellar models
and boundary conditions
(Sec.~\ref{ssec:acchist}).
Our results show that if mass accretion is thermally inefficient,
variation in the accretion history hardly influences
protostellar evolution.
Although isochrones for non-accreting protostars, 
such as those calculated by \citet{dantona94, baraffe98, siess00},
do not necessarily
provide us with correct stellar ages, models with differing
accretion histories nevertheless form a tight sequence in the HRD. Thus
variable accretion histories alone cannot explain the
observed spread of PMS stars in the HR. Moreover, the
errors in absolute ages arise because non-accreting isochrones
are not good descriptions of stars growing with thermally inefficient
mass accretion. They are not a result of variable accretion
histories. However, we note that this does suggest that accreting
isochrones may resolve the problem of systematically larger inferred ages
for high-mass cluster members \citep[e.g.,][]{hillenbrand09,covey10}.

Second, we examine protostellar evolution with different
initial models and thermal efficiencies of mass accretion,
using a constant accretion rate $\dot{M} = 10^{-5}~\msunyr$
(Sec.~\ref{ssec:ini} and \ref{ssec:eff}).
We find that the spread of PMS stars in the HRD that results
from varying the initial radius (or entropy)
or the thermal efficiency is much larger than
the spread that arises from different accretion histories.
We find that a coeval population of stars with significant star-to-star
variation in thermal efficiency or initial radius could conceivably
occupy the entire observed luminosity range for protostars
with effective temperatures $\ga 3500$ K. Thus ages and age spreads
observed in this temperature range may be unreliable. At lower effective
temperature, however, the situation is very different. The only models
we found in our parameter space survey that are capable of producing
false old ages at low $T_{\rm eff}$ are those with purely cold accretion
starting from very small initial radii.
However, we point out that these models require that
$\sim 0.01~M_\odot$ second protostellar cores have radii 
$<0.3~R_\odot$, more than an order of magnitude smaller than
any formed in simulations to date (e.g., MI00), and 
that they can be rendered consistent with observations 
only if such small initial radii are realized
only for stars that end up growing to small final masses. Neither
possibility can be definitively ruled out, but neither is supported
by any current observations or theory either.
If we exclude very small initial radii on these grounds,
we find that all the remaining models indicate that ages and age
spreads inferred from non-accreting isochrones are reliable for
cool stars, at least to the extent that the observationally-determined
luminosities and temperatures are reliable 
\citep[e.g.,][]{dario10a, dario10b}.

In varying the thermal efficiency, we find that models
with only a small amount of hot accretion, e.g., while 
$M_* \le 0.03~\msun$, nonetheless show similar evolution to models 
with entirely hot accretion. Thus,
in order to explain the HRD spread at low-masses with cold
accretion models, the models must begin with small radii and be
thermally inefficient for nearly their entire evolution.
Observational constraints on the thermal efficiency of accretion are somewhat
tenuous. During the earliest stages during which accretion
rates of $10^{-6}-10^{-4}~\msun$ yr$^{-1}$ occur, protostars are too
deeply embedded with too much radiation reprocessing to measure
accretion signatures directly.  Observations of T Tauri stars, for
which accretion rates have declined to $\lesssim 10^{-7}~ \msun$
yr$^{-1}$, suggest that for these rates the accretion column covers 
only $\sim$1-10\% of the stellar surface. This supports a cold accretion 
senario later in the accretion history. However,
observations also find that the covering fraction increases with
accretion rate \citep{gullbring00, ardila00}, suggesting that accretion
may be more thermally
inefficient at early times. Higher mass stars, which experience
higher initial accretion rates,
seem unlikely to avoid hot accretion, while it is more observationally
probable that very low-mass stars experience purely cold accretion. 

Our conclusions are different from those of BCG09, who stressed
the significance of episodic mass accretion for explaining the
observed spread of PMS stars in the HRD.
However, our numerical results are actually consistent with 
theirs. First, BCG09 calculated protostellar evolution using
simple, non-episodic accretion histories (their Fig.~1)
and found that, with thermally efficient accretion ($\alpha \geq 0.2$
in their notation), isochrones for non-accreting protostars 
give the correct ages at $t \gtrsim 1$~Myr.
Disagreement with the isochrones arises only for
thermally inefficient accretion flows ($\alpha = 0$). 
This is consistent with the results shown in our Figure \ref{fig:ri_pms}.
Next, they calculated the evolution with more vigorous, episodic
mass accretion histories (their Fig.~2).
However, the spread they obtain is no broader than that shown in their Figure 1,
indicating that episodic accretion does not increase the HRD spread
beyond what they had already introduced by using varying
initial conditions and thermal efficiencies. Indeed, they state that the 
time-dependence of the accretion rates is not essential for 
their results, and we confirm this finding.\footnote{\citet{baraffe10} 
argue that episodic mass accretion also  
significantly influences the lithium depletion of low-mass stars. However, 
our present work suggests that variation of the initial models
will be more significant than variation of the accretion histories 
for the problem of lithium depletion as well.}
Our results suggest that BCG09 were able to obtain small luminosities
for stars with $T_{\rm eff} \la 3500$ K, and thus claim to reproduce
the observed HRD using a coeval population, because they used
cold accretion starting from initial conditions with entropies far
lower than what current theoretical models predict.
They also did not continue runs with these initial conditions
up to higher masses and effective temperatures.\footnote{For example,
the only cases BCG09 show where stars with $T_{\rm eff} < 3500$ K lie
near the 10 Myr non-accreting isochrone after 1 Myr are their models
A-C. However, they
only use the initial conditions and accretion rates for models A-C to
produce stars up to $0.2$ $\msun$. In comparison, all the models that they
run to masses above $0.5$ $\msun$ have much higher
starting entropies.} Our results 
suggest that, had they done so, the resulting stars would have fallen
well below the locus of observed stars in the HRD, as do our comparable
low initial entropy models.

Finally, we stress that we do not reject the episodic mass accretion
as a possible solution for the ``luminosity problem'' of 
young embedded sources \citep{dunham10, mckee10, offner11}.
Episodic accretion may well occur. It is simply not capable of explaining
the broad spread of optically-visible PMS stars in the HRD.

{\acknowledgements 
We thank Kazu Omukai, Shu-ichiro Inutsuka, Christopher McKee, 
and Kevin Covey for fruitful discussions, and Dawn Peterson for providing us
with her observational data. 
This work was initiated at the meeting ``Multi-phase Interstellar Medium
and Dynamics of Star Formation'' held in Feb. 2010 at Nagoya university. 
T.H. appreciates the support by Fellowship of the Japan
Society for the Promotion of Science for Research Abroad.
M.R.K. acknowledges support from an
Alfred P.\ Sloan Fellowship, from NSF through grants AST-0807739 and
CAREER-0955300, and from NASA through Astrophysics Theory and
Fundamental Physics grant NNX09AK31G and through a Spitzer Space Telescope
Cycle 5 Theoretical Research Program grant. 
S.S.R.O acknowledges support from NSF grant AST-0901055.
} 

\clearpage

\bibliography{biblio}
\bibliographystyle{apj}

\end{document}